\def\lesssim{{_ <\atop{^\sim}}}

\def\ap3m{AP$^3$M}
\def\LCDM{$\Lambda$CDM}
\def\LWDM{$\Lambda$WDM}
\def\hkpc{$h^{-1}{\ }{\rm kpc}$}
\def\hMpc{$h^{-1}{\ }{\rm Mpc}$}
\def\hMsun{$h^{-1}{\ }{\rm M_{\odot}}$}
\def\kms{${\rm{\ }km{\ }s^{-1}}$}
\def\nbody{$N$-body}
\def\c15{$c_{\rm 1/5}$}
\def\rvir{$r_{\rm vir}$}

\def\ea{et~al.~}                            

\def\lesssim{\mathrel{\hbox{\rlap{\hbox{\lower4pt\hbox{$\sim$}}}\hbox{$<$}}}}
\def\gtrsim{\mathrel{\hbox{\rlap{\hbox{\lower4pt\hbox{$\sim$}}}\hbox{$>$}}}}

\newcommand{\AAA}[3]    {\mbox{A\&A~\textbf{#1},~#2~(#3)}}

\newcommand{\ApJ}[3]    {\mbox{ApJ~\textbf{#1},~#2~(#3)}}

\newcommand{\ApJL}[3]   {\mbox{ApJ~Lett.~\textbf{#1},~#2~(#3)}}

\newcommand{\AJ}[3]     {\mbox{Astron.~J.~\textbf{#1},~#2~(#3)}}
\newcommand{\MNRAS}[3]  {\mbox{MNRAS~\textbf{#1},~#2~(#3)}}
\newcommand{\Nature}[3] {\mbox{Nature~\textbf{#1},~#2~(#3)}}

\newcommand{\Science}[3]{\mbox{Science~\textbf{#1},~#2~(#3)}}

\newcommand{\astroph}[1]{\mbox{\texttt{astro-ph/#1}}}

\documentstyle[epsfig]{mn}

\begin{document}

\title{Merger histories in WDM structure formation scenarios}

\author[Knebe A. et al.]
       {Alexander Knebe$^{1}$ Julien E. G. Devriendt$^{2}$, 
        Asim Mahmood$^{2}$ and Joseph Silk$^{2}$\\        
       {$^1$Theoretical Physics, Keble Road, Oxford OX1 3NP, UK}\\
       {$^2$Astrophysics, Keble Road, Oxford, OX1 3RH, UK}}

\date{Received ...; accepted ...}

\maketitle

\begin{abstract}
Observations on galactic scales seem to be in contradiction with
recent high resolution \nbody\ simulations. This so-called cold dark matter
(CDM) crisis has been addressed in several ways, ranging from 
a change in fundamental physics by introducing
self-interacting cold dark matter particles to a tuning of complex astrophysical processes
such as global and/or local feedback. 
All these efforts attempt to soften density profiles and reduce the
abundance of satellites in simulated galaxy halos. 
In this paper, we explore a somewhat different approach which consists of
filtering the dark matter power spectrum on small scales, thereby 
altering the formation history of low mass objects. The physical motivation 
for damping these fluctuations lies in the possibility that the dark matter particles
 have a different nature {\em i.e.} are warm (WDM) rather than cold.
We show that this leads to some interesting new results
in terms of the merger history and large-scale distribution
of low mass halos, as compared to the standard CDM scenario.
However, WDM does not appear to be the ultimate solution, in the sense 
that it is not able to fully solve the CDM crisis,
even though one of the main drawbacks, namely the abundance of satellites,
can be remedied. Indeed, the cuspiness of the halo profiles still
persists, at all redshifts, and for all halos and sub-halos that we investigated. 
 Despite the persistence of the 
cuspiness problem of DM halos, WDM seems to be still worth  taking seriously,
as it  alleviates the problems
of over-abundant sub-structures in galactic halos and possibly the {\em lack} of
angular momentum of simulated disk galaxies.
WDM also  lessens the need to invoke  strong feedback 
to solve these problems, and may provide a natural explanation of
the clustering properties and ages of dwarfs.

\end{abstract}


\section{Introduction}

CDM models have been very successful in reproducing the large scale
structure properties of the universe ({\em e.g.} Bahcall et al.,
1999). However, they have lately been facing a state of crisis because
of apparent discrepancies between high resolution $N$-body simulations
and observations on galaxy scales. One can divide these problems into
two categories: the cuspiness of typical L$_\star$ galaxy halos on the
one hand (cf. Moore~\ea 1999a), and the dearth of dark matter
satellites in these very same halos on the other (Klypin et al. 1999,
Moore et al. 1999b). As a matter of fact, high resolution observations
of galaxy rotation curves seem to be quite incompatible with steep
dark matter cores (de Blok \ea 2001, even though Van den Bosch et
al. (2000) have challenged this idea), and the number of observed
satellites of the Milky-Way is about an order of magnitude smaller
than that measured in cold dark matter
\nbody\ simulations (Klypin et al. 1999). Furthermore, micro-lensing experiments 
toward the galactic centre lead to the conclusion that cuspy dark
matter profiles might yield too much mass inside the solar radius
(Binney, Bissantz~\& Gerhard 2000).

These problems have triggered a series of papers devoted to changing
the properties of the dark matter itself and making it
self-interacting ({\em e.g.} Spergel and Steinhardt, 2000; Bento et
al., 2000).  However, it is still unclear that we need such dramatic
changes to reconcile theory and observations as, for instance, massive
black holes in the centre of galaxies could alleviate/solve the cusp
problem (Merritt~\& Cruz 2001), and re-ionisation of the universe
could get rid of a significant fraction of visible low mass satellites
(Chiu~\ea 2001).  Bearing these possibilities in mind, we adopt a
rather conservative view which has the advantage of being well
motivated on the particle physics side.  To be more specific, we note
that, in principle, active neutrino oscillations can naturally produce
sterile neutrinos with masses of order 1 keV (Dolgov and Hansen,
2000). Also, such warm particles would preserve a CDM-like scenario of
large scale structures, which is known to be quite successful
(Colombi~\ea 1996; Bahcall~\ea 1999).

The outline of the paper is as follows.  We first start with a brief
description of the initial power spectra in Section~\ref{WDMpower},
then move on to the simulations themselves in Section~\ref{Nbody}. In
Section~\ref{Analysis} we analyze the output from our numerical
simulations mainly focusing on the merger history in the new WDM
models. We finally summarize our results and conclude in
Section~\ref{Conclusions}.

\section{Warm Dark Matter Power Spectra} \label{WDMpower}

In agreement with the combined observations of the cosmic microwave
background anisotropies on sub-degree scales by BOOMERanG (de
Bernardis~\ea 2000) and MAXIMA (Balbi~\ea 2000), and of high-redshift
supernovas (Riess~\ea 1998; Schmidt~\ea 1998; Perlmutter~\ea 1999), we
have chosen a flat universe model with a cosmological constant. More
specifically, values for the cosmological parameters in all the
simulations presented here are: $\Omega_0 =$ 0.33, $\Omega_\Lambda =$
0.67, $H_0 =$ 67 \kms Mpc$^{-1}$, $\sigma_8 =$ 0.88. We note that in
the case of a CDM model, these values also account correctly for the
large-scale structure properties of the universe, such as the
evolution of cluster abundances (Eke~\ea 1998) and the distribution of
galaxies (Benson~\ea 2001).

The only difference in the power spectra of the three simulations
presented here comes from the damping of small-scale density
fluctuations due to relativistic free-streaming in the WDM
models. More specifically, we have chosen the free-streaming scales, R$_f$,
to be 0.2 and 0.1 \hMpc\ for our WDM simulations,
corresponding to smoothing scales
--- defined as the comoving half-wavelength of the mode for which the
amplitude of the linear density fluctuation is divided by two --- of 
0.72 and 0.33 \hMpc\ respectively and therefore to
masses of 0.5 keV and 1.0 keV for the respective warmons (cf. Eq.~(1)
of Bode, Ostriker and Turok (2000)).  The values are summarized in
Table~\ref{WDMparameter}.

\begin{table}
\caption{Parameter of the WDM models.}
\label{WDMparameter}
\begin{tabular}{lll} \hline \hline
               & \LWDM 1   & \LWDM 2     \\ \hline
 $R_f$         & 0.1 \hMpc & 0.2 \hMpc    \\ \hline
 $m_{\rm WDM}$ & 1.0   keV & 0.5 keV  \\ \hline \hline
\end{tabular}
\end{table}

Following Bardeen~\ea (1986), the WDM power spectra, P$_{\rm WDM}$, 
are then obtained by multiplying the CDM power spectrum, P$_{\rm CDM}$, by 
the filter function T$^2_{\rm WDM}$, where:

\begin{equation}
 T_{\rm WDM} (k) = \exp \left( -\frac{k R_f}{2} - \frac{(k R_f)^2}{2} \right).
\end{equation}

\begin{figure}
      \centerline{\psfig{file=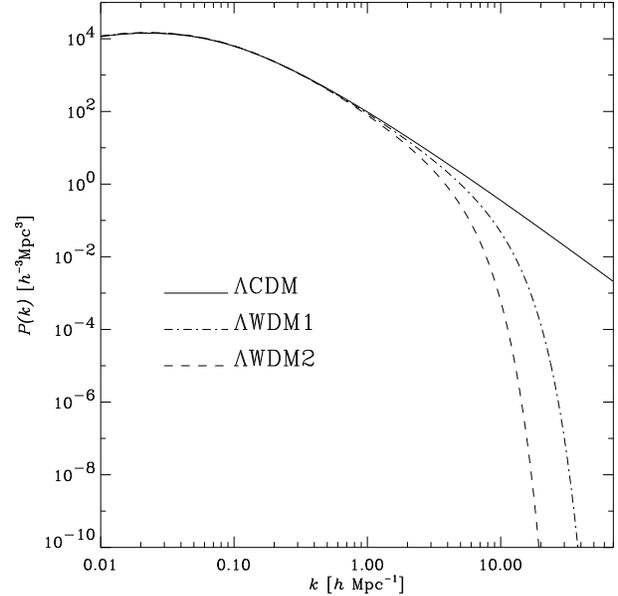,width=\hsize}}
      \caption{Input power spectra.}
      \label{PkInput}
    \end{figure}

\noindent
Fig.~\ref{PkInput} shows the initial power spectra which are used as
an input for generating the initial conditions of our three different
dark matter simulations analyzed in Section~\ref{Analysis}.

Note that these choices for the masses of the warmons are compatible
with observed properties of the Lyman-$\alpha$ forest in high-redshift
quasar spectra and with the fact that a minimal fraction of baryons
has to have already collapsed by z $\approx$ 6 if the universe is to
be re-ionized at high-redshift.
As a matter of fact, this latter constraint leads to a lower limit for the 
warmon mass $m_{\rm WDM} \approx 1.2$ keV, which can go down to 
$m_{\rm WDM} \approx 750$ eV in the somewhat extreme case where the 
ionizing-photon production efficiency is ten times greater at high 
redshifts (Barkana, Haiman~\& Ostriker 2001). On the other hand, such a low
value for the lower limit of the warmon mass has also been derived from 
analysis of artificial Lyman-$\alpha$-spectra extracted from numerical 
hydrodynamics simulations (Narayanan~\ea 2000). 

We have not explicitly
assigned non-zero initial thermal velocities for particles in the
simulations, because as shown by Hogan and Dalcanton (2000) from phase
space density arguments, even if present, they would be too small to
be relevant on dwarf galaxy halo scales for the masses of the warmons
under investigation here.  We note that this conclusion is also
supported by the recent simulations of Avila-Reese~\ea
(2000) and Bode~\ea (2000) who have included such
velocities in their WDM \nbody-simulations.

\section{The $N$-body Simulations} \label{Nbody}
All three simulations were carried out using the multiple-mass
Adaptive Refinement Tree code (ART; Kravtsov, Klypin \& Khokhlov
1997).

The ART code achieves high force resolution by decreasing the size of
grid cells in all high-density regions, using an automated refinement
algorithm.  These refinements are recursive, which means that refined
regions can also be refined, each subsequent refinement having cells
that are half the size of the cells in the previous level.  This
creates a hierarchy of refinement meshes of different resolution
focusing on regions of high density where high spatial force
resolution is needed.  The present version of the code uses multiple
time steps on different refinement levels, as opposed to the constant
time stepping in the original version of the code. The multiple time
stepping scheme is described in detail in Kravtsov et al. (1998).

In addition to these features, the latest version of the ART code also
allows the usage of multiple-masses. We started by placing $512^3$
particles into the simulation volume according to the input power
spectra as given by Fig.~\ref{PkInput} and then using the Zeldovich
approximation. All these particles were then collapsed in packets of
eight until only $128^3$ particles were left. Therefore, in order to
re-simulate a region of interest with higher mass resolution we would
just need to 'unpack' the high-mass particles within that region which
automatically adds the correct high frequency waves to the
simulation. The analysis of such runs will be deferred to a companion
paper. Nonetheless, as we reduced our initial number of $512^3$
particles only by a factor 64 to obtain $128^3$ particles, we already
reach the interesting mass resolution of $m_p \sim 7 \cdot
10^{8}$\hMsun\ in the "low-resolution" runs that we discuss in this
paper.

The box size was chosen to be $25$\hMpc\ on a side and the
distribution of particles was evolved from a redshift $z=35$ to $z=0$
in 1000 integration steps, reaching refinement level~5 in all three
runs. Because of the multiple time stepping, this corresponds to 32000
steps on the finest grid. As we started with a regular grid of $512^3$
grid cells covering the whole computational volume we reached a force
resolution of 3\hkpc. All these parameters are summarized in
Table~\ref{SimuParam}.

\begin{table}
\caption{Specifications of the numerical simulations.}
\label{SimuParam}
 \begin{tabular}{ll} 
  Simulation Parameter \\\hline
  box size             & 25 \hMpc \\
  initial redshift     & $z_i=30$ \\
  number of particles  & $128^3$ \\
  integration steps    & 1000 \\
  force resolution     & 3 \hkpc \\
  mass resolution      &  $6.89 \cdot 10^{8}$\hMsun \\

 \end{tabular}
\end{table}

We output snapshots of the simulation at the following redshifts
$z=5,4,3,2,1,0.8,0.6,0.5,0.4,0.3,0.2,0.1,0.0$. Then, we identify
galaxy halos and their substructure content using a standard
friends-of-friends group finding algorithm (FOF, Davis \ea 1985) as
well as the more sophisticated Bound-Density-Maxima code (BDM,
Klypin~\& Holtzman 1997). The BDM algorithm finds the positions of
local density maxima smoothed out on a certain scale of interest and
uses these maxima as centres for calculating {\em e.g.} density
profiles in a pre-selected number of radially placed bins. While doing
so, the program decides whether particles are bound or unbound and
removes the latter ones from the halo in an iterative way.

The final halo catalogues (either FOF or BDM) constitute the building blocks of
the analysis described in the following section.

\section{Analysis} \label{Analysis}

\subsection{Visualisation} \label{visualisation}
\begin{figure}
      \centerline{\psfig{file=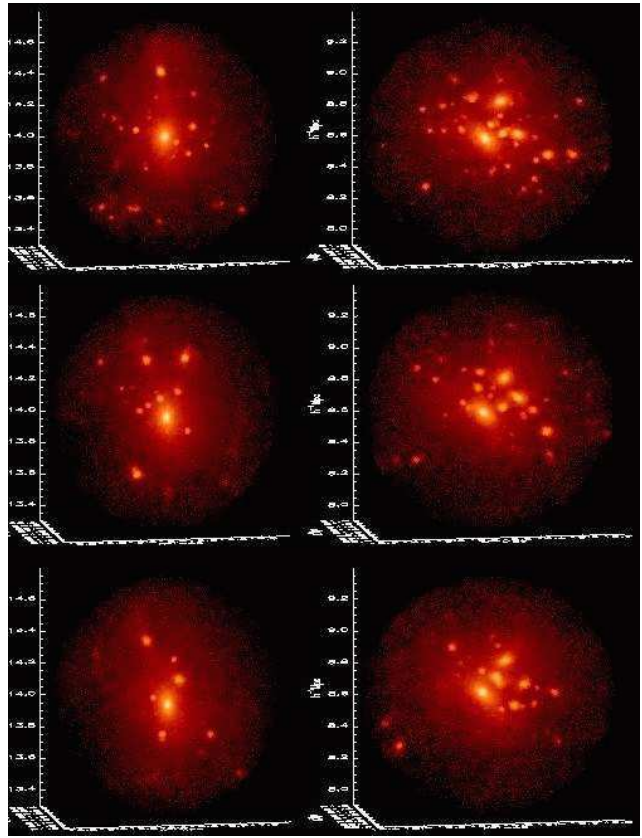,width=\hsize}}
      \caption{Grey scaled density field of the two most massive
               galactic halos identified at z=0. The left panel shows 
               halo~\#1 of mass $9.3\cdot 10^{13}$\hMsun\ (about 
               135000 particles) and the right panel shows halo~\#2 
               weighing $8.2\cdot 10^{13}$\hMsun\ (ca. 120000 particles).}
      \label{haloes}
    \end{figure}

The first thing we present is the distribution of dark matter
particles in a sphere of radius 750 \hkpc\ centred around the two
most massive halos in all three simulations. A grey-scaled
view of the density fields is given in Fig.~\ref{haloes}, where
the density at each particle position was calculated on a 
$256^3$ cell grid.

There are two things to notice in this plot: 
\begin{itemize}
\item we clearly see a decrease of substructure 
      (fewer sub-halos) when moving from CDM to WDM scenarios, 
\item only the left halo (from now
      on called halo~\#1) shows a well defined central core. For the
      right halo (named halo~\#2) is it difficult to locate the centre not
      only by eye, but also using the BDM method: this halo is in fact
      undergoing a merger event, and therefore has a quite different mass
      build-up history than halo~\#1. 
\end{itemize}

To sum up, we are left with the fortunate situation where we can study
the merger history of two well resolved halos (more than $10^5$
particles within the region shown in Fig.~\ref{haloes}), one being
more relaxed than the other in the CDM structure formation scenario as
well as in the new WDM models.

\subsection{Power Spectrum Evolution}

The first thing to be checked quantitatively is the evolution of the
dark matter power spectrum $P(k)$ to see if the filtering scale has
left an imprint in today's power spectrum.  Knebe, Islam~\& Silk
(2001) have recently shown that Gaussian features in $P(k)$ are washed
out by non-linear effects and it is interesting to see whether or not
this also happens with a filtered power spectrum.

\begin{figure}
      \centerline{\psfig{file=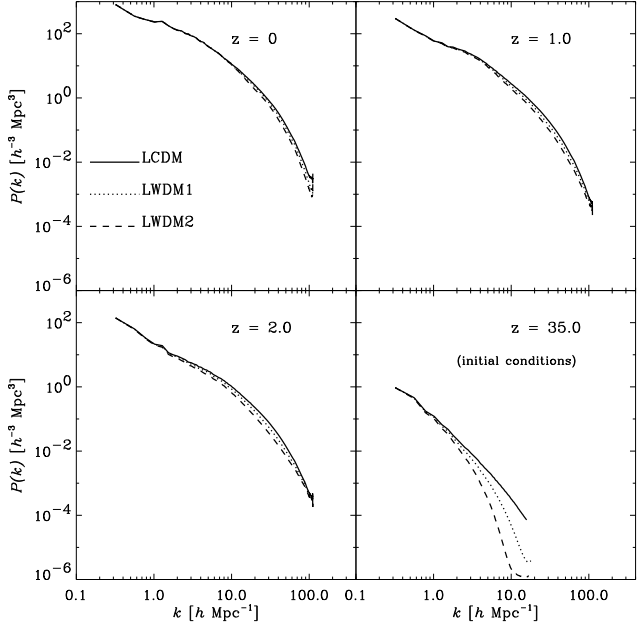,width=\hsize}}
      \caption{Power spectrum evolution.}
      \label{power}
    \end{figure}

Fig.~\ref{power} clearly proves that non-linear evolution of the power
spectrum boosts the power on small scales so dramatically that we find
almost no trace of the 'cut-off' after a while. Already at a redshift
of $z=2$, the WDM spectra has almost converged to the unfiltered CDM
spectrum. Therefore, if bias was linear on these scales, we would not expect 
to find any signature of WDM-like spectra, even in the latest galaxy surveys 
such as the PSC$z$ (Hamilton~\& Tegmark 2000).

\subsection{Mass Function of Halos} \label{mass}
The most basic property of a dark matter halo is its mass. In spite
of this, mass provides a wealth of information about the formation of
structures, especially when computing the (cumulative) distribution of
objects $n(>M)$ with mass $M$.  As we are expecting the formation of
small halos to be damped when filtering the input power spectrum, it
indeed seems logical to begin with the study of gravitationally bound
objects' abundances.

\begin{figure}
      \centerline{\psfig{file=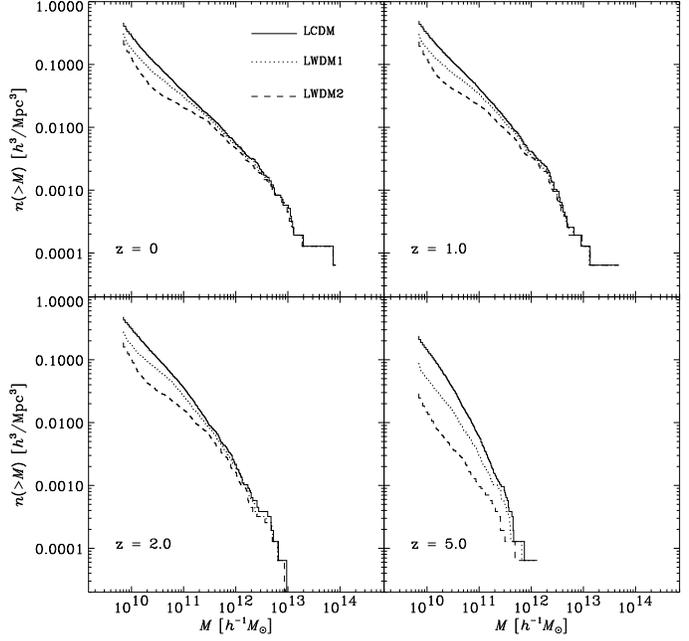,width=\hsize}}
      \caption{Evolution of mass function.}
      \label{massfunc}
    \end{figure}

Therefore, in Fig.~\ref{massfunc} we show halo mass functions in our 
three dark matter simulations at redshifts $z=0$, $z=1$, $z=2$, 
and $z=5$.  We can
clearly see how structures develop in the bottom-up fashion in all
three models, although both WDM scenarios exhibit a noticeably
different behaviour at the low-mass end. As a matter of fact, we
observe in these simulations far less small mass objects than in the CDM
simulation, with the
deviation being more pronounced and spanning a wider mass range at earlier
times. This can be explained by the fact that the formation of low
mass objects is somewhat hampered at early times (z $>$ 2 ) by the
cut-off in the power spectrum.

On the other hand, we also notice that the mass function in our WDM
simulations becomes steeper than in the CDM simulation at the low mass
end ($M < 3 \cdot 10^{10}$\hMsun) and for smaller redshifts.  This
phenomenon was also observed by Bode, Ostriker~\& Turok (2000) and can
be explained by formation of ``roughly below the free-streaming mass''
halos (in our case $1.3 \cdot 10^10$\hMsun\ for
\LWDM1 and $2.7 \cdot 10^10$ \hMsun\ for \LWDM2), mainly through
``pancake'' or filament fragmentation. We point out that this is a
rather crude estimate as it is clear from Fig.~\ref{massfunc}, that it
affects halos which are several times more massive than these
limits.  Furthermore, as noted by these authors, as these halos are
preferably formed along filaments in a ``top-down'' scenario, their
spatial distributions and epochs of formation contrast with those of a
CDM model. This seems to be in better agreement with the clustering
measurements of the local population and estimated ages of dwarfs
(Peebles 2000; Metcalfe et al. 2000) to which these halos correspond.

\begin{figure}
      \centerline{\psfig{file=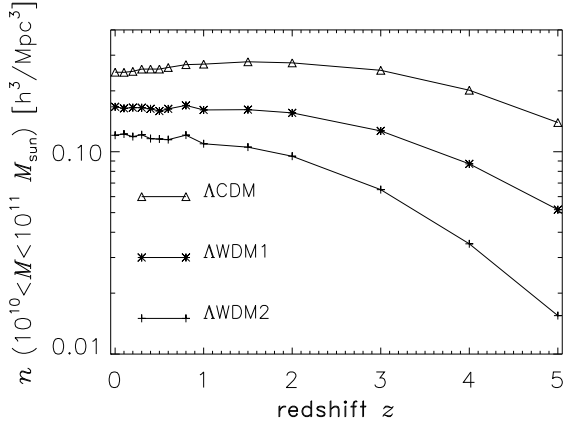,width=\hsize}}
      \caption{Evolution of halo abundance for particle groups with 
               mass $M$ in the range [$10^{10}$\hMsun,$10^{11}$\hMsun].}
      \label{abundance}
    \end{figure}

In addition, we plot the number density evolution of objects in the
mass range $M\in$[$10^{10}$\hMsun,$10^{11}$\hMsun] in
Fig.~\ref{abundance}. From this figure, one can clearly see that there
are fewer small mass halos at high redshifts in WDM simulations.
As the difference with the number of CDM halos of the same mass is
less important at low redshift, we can therefore conclude that their
average formation epoch is much more recent.  As previously mentioned,
the majority of these objects do not follow the standard CDM
``bottom-up'' formation process but are rather formed in a
``top-down'' fashion.  We will come back to this statement in the next
Section, when investigating the dependence of the position of the
objects on the environment.

   \begin{figure} \centerline{\psfig{file=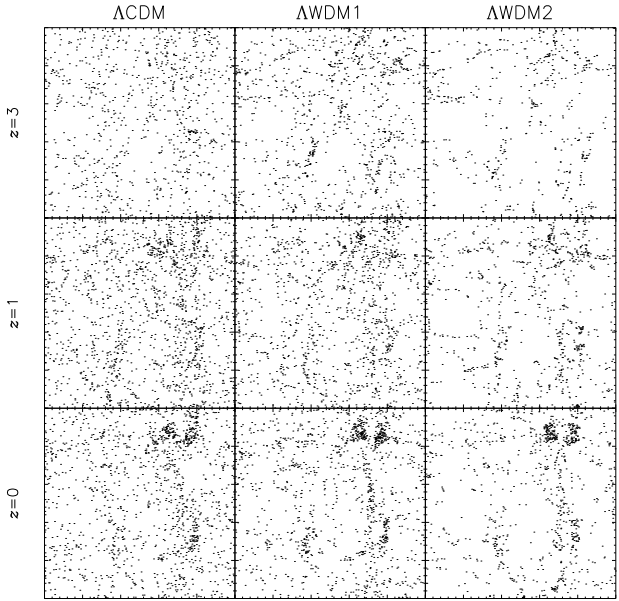,width=\hsize}}
    \caption{Position of halos in the mass range
    $M\in$[$10^{10}$\hMsun,$10^{11}$\hMsun].}  \label{HaloR}
    \end{figure}

   \begin{figure} 
    \centerline{\psfig{file=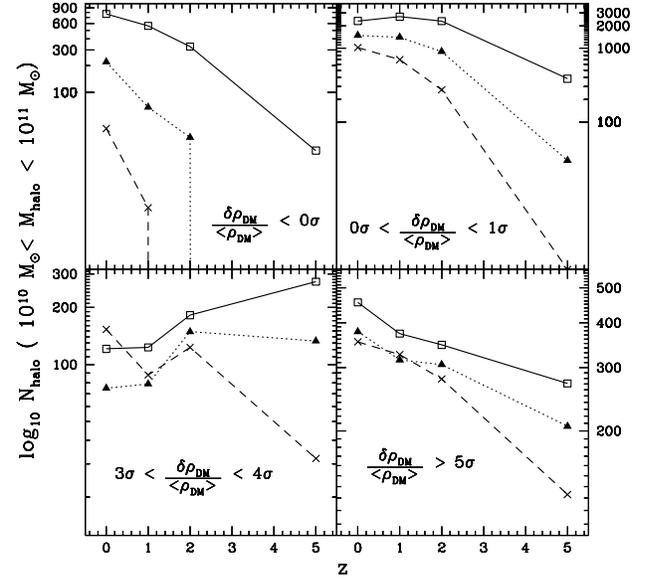,width=\hsize}}
    \caption{Time evolution of the number of halos in the mass range
             $M\in$[$10^{10}$\hMsun,$10^{11}$\hMsun] located within
             a region of a given density contrast.}  
    \label{halodc}
   \end{figure}
\subsection{Halo Positions} \label{positions}
In the previous Section~\ref{mass} (Fig.~\ref{abundance}), we have
seen that the evolution of the number density of objects in the mass
range $M\in$[$10^{10}$\hMsun,$10^{11}$\hMsun] differs significantly
from one model to another. At a redshift of $z=5.0$ the abundance of
these small mass halos is about a factor of 10 larger in the
\LCDM\ simulation than in the \LWDM2 simulation, whereas the ratio drops 
to about 2 at $z=0.0$.  We claim that in WDM structure formation
scenarios, those objects are primarily formed in high density regions
by pancake and filament fragmentation. To back up this statement we
first plot in Fig.~\ref{HaloR} the projected positions of the halos
within the aforementioned mass range onto the $x-y$-plane of the box.  We
can now see that in the \LWDM\ simulations, low mass halos are indeed
tracing the filamentary structure, in contrast to the \LCDM\
simulation where they are much more void-filling. We mention that for
clarity, at a given redshift, all plots contain the same number of
halos, which means that we have randomly chosen a subset of objects
for plotting the positions of halos in the \LCDM\ and \LWDM1\ models,
while using the complete data set for \LWDM2.  As a complement, in
Fig~\ref{halodc}, we plot the redshift evolution of the number of low
mass halos located in regions of various density contrasts. The density
field was computed on a comoving grid composed of $25^3$ cells, using a 
nearest grid point method to assign particles to the grid. This means that 
each grid cell is exactly 1 \hMpc\ on a side. 
We then repeat exactly the same procedure, assigning halos with masses 
between $10^{10}$\hMsun and $10^{11}$\hMsun to their nearest grid point.
The standard deviations, $\sigma$, for the density field, range from 
being about 9 times greater than the mean density at $z = 0$, when the 
particle distribution is highly clustered, to 0.7 times lower than the mean 
density at $z = 5$, when the particle distribution is quite smooth. 
This means that if the density contrast in one region at $z = 0$ is greater 
than 2 $\sigma$, this region is about 18 times denser than the average
density of the universe at this redshift. Notice that the standard deviations
from the mean density are about the same in the CDM and the WDM simulations,
and for all redshifts (within 10 \%). 
It is obvious from this figure that halos in this mass range are more
strongly suppressed in low density regions and/or at high redshifts in
\LWDM\ cosmologies.  On the other hand, in high density regions and at
fairly low redshifts, things are quite different: one sees more halos
in regions with density contrast between 3 and 4 $\sigma$ (bottom
left panel of Figure~\ref{halodc}) at $z = 0$ in simulation
\LWDM2 than in simulation \LCDM! This is a non ambiguous signature of late 
``bottom-up'' formation of low mass halos in these regions.  However,
as we are speaking of objects containing only about 20-200 particles
we cannot exclude this effect to be due to Poisson fluctuations at the
100\% level.  But if we check the representation of the initial power
spectrum at $z=35$ (cf. Fig.~\ref{PkInput}) we confirm to sample waves
out to at least wavenumber $k \approx 15 h {\rm Mpc}^{-1}$ in all
three models. This corresponds to a mass scale of roughly $3 \cdot
10^{10}$\hMsun\ and hence Poisson noise probably only affects even
smaller objects.

   \begin{figure}
      \centerline{\psfig{file=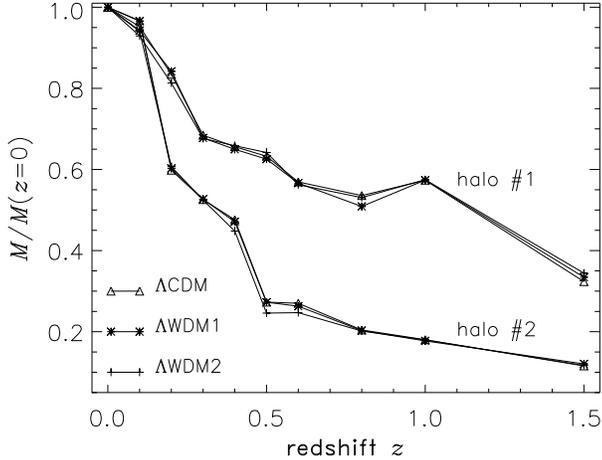,width=\hsize}}
      \caption{Evolution of halo mass for halo~\#1 and~\#2 in all
               three models.}
      \label{MassHistory}
    \end{figure}

\subsection{Merger Histories}

As we already mentioned in Section~\ref{visualisation} we expect the
merger histories of the two halos~\#1 and~\#2 to be
different. Therefore we investigate how the mass of these halos (normalized
to the present value) evolves as a function of redshift, and present the results
for both halos and all three dark matter models in Fig.~\ref{MassHistory}.
Tracing back the merging history of each halo was performed as follows: we first
identify all particles belonging to the halo at redshift
$z=0$ and locate these particles at redshift $z=0.1$.  The halo which
now contains the majority of these particles is tagged as the main
progenitor and its mass is stored. We repeat the same procedure
iteratively out to redshift $z=5.0$. The results are only plotted to
redshift $z=1.5$ as the formation time of a halo is normally defined
to be the redshift where it weighs half of its present mass.
This is about $z=1.0$ for halo~\#1 and between $z=0.3$ and $z=0.4$ for
halo~\#2.

The figure clearly indicates that two major mergers
occurred during the lifetime of halo~\#2; one around $z=0.5$
and a more recent one between $z=0.2$ and $z=0.1$. These
epochs will be particularly interesting when we will analyse
the evolution of the density profile of the halo to
assess the impact of merger events upon it. Halo~\#1, on the other hand,
builds up its mass mainly via accretion for we do not observe
such rapid changes in its merging history.

However, the most important thing to notice in Fig.~\ref{MassHistory}
is that there are almost {\em no} differences in the merger histories
of CDM and WDM halos.  This statement is valid up to $z=5$ and for
smaller particle groups not shown in Fig.~\ref{MassHistory}; we
followed the merger histories of several objects less resolved than
our halos~\#1 and~\#2 and did not find any significant discrepancy
between CDM and WDM models. This is rather remarkable as there 
definitely is a smaller number of satellites orbiting those halos at $z=0$ in
the WDM scenario, as will be shown in
Section~\ref{satellites}. Furthermore, the final masses of halos \#1 and \#2
are very close to one another, in all three dark matter simulations,
which indicates that these halos have probably been built up from a
similar population of small mass objects.  In
Section~\ref{satellites}, we actually compute the number of satellites in 
both halos at various redshifts. It then becomes clear that there are indeed
comparable numbers of such small mass halos orbiting in halo~\#1
and~\#2 at earlier times. One then has to advocate that in the
WDM models, such satellites are disrupted and have their mass
redistributed over the whole host halo. Colin, Avila-Reese and
Valenzuela~(2000) showed that this indeed is the case, as the
number of satellites is nearly identical in CDM and WDM at $z=1$ in their 
simulations: the "suppression" of small scale structures in massive halos has
to result from a disruption of sub-halos rather than from a different
merger history.

\subsection{Satellite Abundances} \label{satellites}
\begin{figure}
      \centerline{\psfig{file=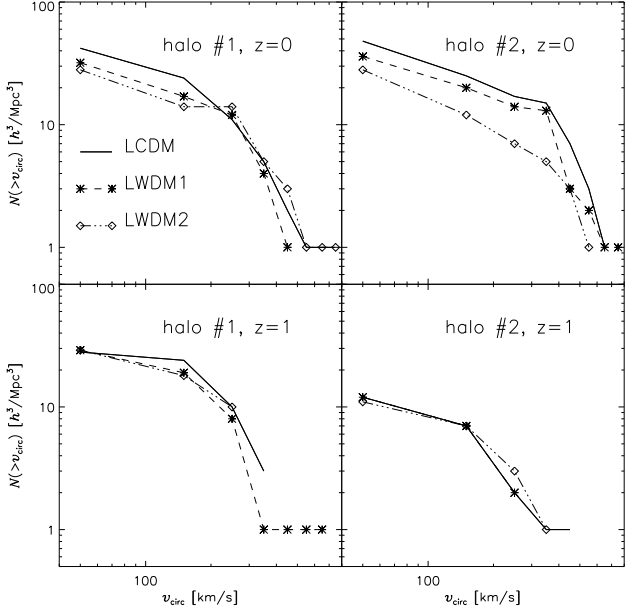,width=\hsize}}
      \caption{Cumulative circular velocity distribution
               host halo~\#1 and~\#2, respectively. Only halos
               more massive than $10^{10}$\hMsun\ are taken
               into account.}
      \label{Nvcirc}
    \end{figure}

One of the problems with CDM models is that high resolution simulations tend
to show an order of magnitude more substructure in galaxy size haloes than is 
actually observed (Klypin~\ea
1999, Moore~\ea 1999b). One should then require of a viable alternative model 
that it naturally reduces the abundance of such satellite halos.
As the results are generally presented as
cumulative velocity distributions for the satellites
$N(>v_{\rm circ})$, we have calculated this function for both our
halos at redshifts $z=0$ and $z=1$. Results are shown in Fig.~\ref{Nvcirc}.

\begin{table}
\caption{Number of satellite galaxies within halo \#1 and~\#2.
         The mass cuts are measured in \hMsun.}
\label{nsatellites}
 \begin{tabular}{l||cc|cc} 
  \multicolumn{5}{l}{\large $z=0.0$} \\ \hline
  & \multicolumn{2}{c}{halo \#1} & \multicolumn{2}{c}{halo \#2} \\ \hline
  &          $M>10^{10}$ & $M>10^{11}$ & $M>10^{10}$ & $M>10^{11}$ \\ \hline \hline
 \LCDM   &      42      &     26      &    48      &  28 \\
 \LWDM 1 &      32      &     19      &    36      &  24  \\
 \LWDM 2 &      28      &     17      &    28      &  18  \\
\\
\\
  \multicolumn{5}{l}{\large $z=1.0$} \\ \hline
  & \multicolumn{2}{c}{halo \#1} & \multicolumn{2}{c}{halo \#2} \\ \hline
  &          $M>10^{10}$ & $M>10^{11}$ & $M>10^{10}$ & $M>10^{11}$ \\ \hline \hline
 \LCDM   &      29      &     20      &    12      &  8 \\
 \LWDM 1 &      29      &     19      &    12      &  9  \\
 \LWDM 2 &      29      &     19      &    11      &  6  \\

 \end{tabular}
\end{table}

To generate our data, we have identified all sub-halos using the BDM
code and used the maximum of the circular velocity curve to define
$v_{\rm max}$. We counted all gravitationally bound particles groups
within a sphere of radius $R=1$\hMpc\ around the centre of the halo
for $z=0$ and $R=0.5$\hMpc\ for $z=1$, respectively. The total number
of satellites found are summarised in Table~\ref{nsatellites} for two
mass cuts $M_{\rm satellite} > 10^{10}$\hMsun\ and $M_{\rm satellite}
> 10^{11}$\hMsun.  Due to the finite number of particles present in
our simulated box, we are not able to resolve halos with maximum
circular velocities below 50\kms. We are aware that the actual
discrepancy between numerical simulation and observations as pointed
out by Klypin~\ea (1999) and Moore~\ea (1999b) appears to be at
velocities smaller than this, but we can safely say that our data
agrees with the results of Colin, Avila-Reese~\& Valenzuela (2000)
when extrapolating their distribution function to our velocity range.
Moreover, at $z = 0$, we already observe a drop in the number density of
satellites with $v_{\rm max} \approx 50$\kms
(cf. Fig.~\ref{Nvcirc}), which is all the more pronounced when the mass 
of the warm dark matter particle $m_{\rm WDM}$ is low.

We have already seen that the masses and mass histories of the two halos are
 quasi identical irrespective of the model, but we count fewer satellites in
the WDM scenarios. How can this happen? One needs to go back to $z=1$
to understand the origin of this puzzle.  At this redshift, we see (both in
Fig.~\ref{Nvcirc} and Table~\ref{nsatellites}) that the number of
sub-halos is indistinguishable between our models; even the
distribution functions $N(>v_{\rm circ})$ can be superimposed.  We
therefore conclude that the same number of satellites are in place by
$z=1$ in both the WDM and CDM simulations, but that for the former, they
are more easily disrupted during the relaxation process of the host
halo. This is also shown in Fig.~\ref{halodc}, where we see that in
regions of very high density contrast ($> 5 \sigma$), and at high
redshifts, the number of small mass halos is about the same, in all
three simulations. This therefore provides a natural explanation as to
why merger histories are so similar for massive halos, in WDM and CDM
simulations. However, one should bear in mind that the equality of the
number of satellites for e.g. halo~\#1 at $z=0$ and $z=1$ in the 
\LWDM2 simulation does not
mean that these sub-halos have survived unaltered. These numbers are the 
result of a competition between
continuous destruction and accretion of satellites during the merging history
of each (host) halo. Therefore, numbers in Table~\ref{nsatellites}
actually indicate that the rates at which these two processes (destruction 
and accretion) occur, become almost identical for halos formed
in the \LWDM2 structure formation scenario.

   \begin{figure}
      \centerline{\psfig{file=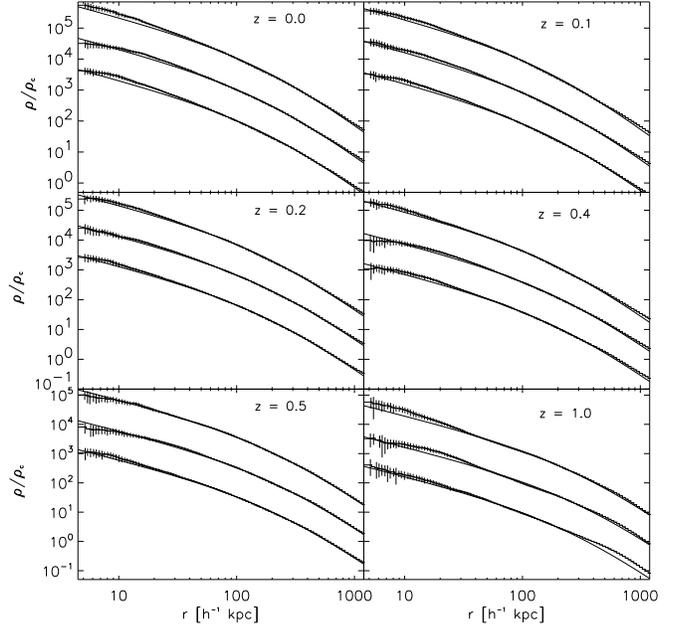,width=\hsize}}
      \caption{Density profiles for halo~\#1 in all three simulations 
               at redshift $z=0$. The curves in each panel are
               shifted by factors of ten downwards for clarity and
               correspond to (from top to bottom): \LCDM, \LWDM1, and
               \LWDM2.}
      \label{DensProfile1}
    \end{figure}

   \begin{figure}
      \centerline{\psfig{file=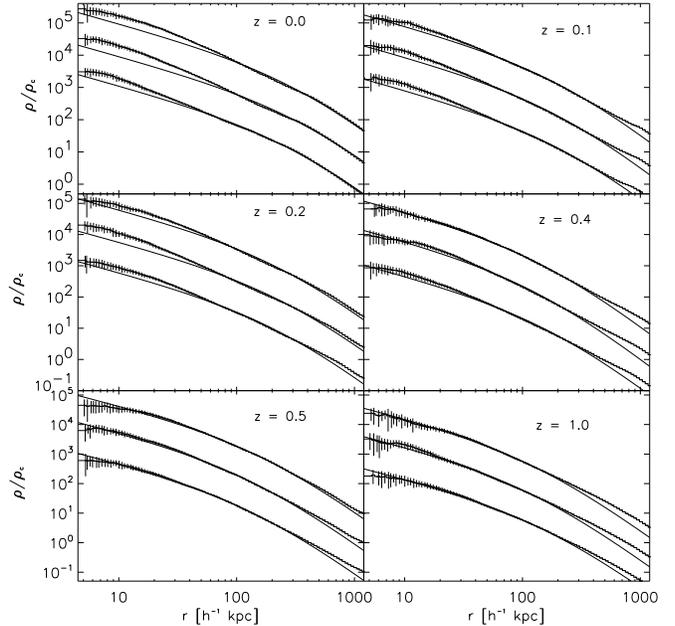,width=\hsize}}
      \caption{Density profiles for halo~\#2 in all three simulations 
               at redshift $z=0$. The curves in each panel are
               shifted by factors of ten downwards for clarity and
               correspond to (from top to bottom): \LCDM, \LWDM1, and
               \LWDM2.}
      \label{DensProfile2}
    \end{figure}

\subsection{Density Profiles}

We now turn to the radial distribution of mass in the halos by
measuring the density profiles of the two most massive halos~\#1
and~\#2 for a series of redshifts between $z=1.0$ and $z=0.0$. We fit
the data to profiles of the form proposed by Navarro, Frenk~\& White
(1997):

\begin{equation}
\frac{\rho(r)}{\rho_c} = \frac{\delta_c}{r/r_s (1+r/r_s)^2} \ ,
\end{equation}

\noindent
where $\rho_c$ is the critical density, $\delta_c$ a characteristic
(dimensionless) over-density and $r_s$ a scale radius. We define the
virial radius \rvir\ of the halo to be the radius where 
$\rho(r_{\rm vir})/\rho_c = 200$. This radius is then
used to calculate the concentration parameter $c_{1/5}$, which is 
defined as the ratio of the virial radius \rvir\ and a scale radius 
$r_{1/5}$, where 1/5 of the virial mass is contained:

\begin{equation} \label{NFWfit}
  c_{1/5}  =  r_{\rm vir}/r_{1/5} \ .
\end{equation}

\noindent

\begin{figure}
      \centerline{\psfig{file=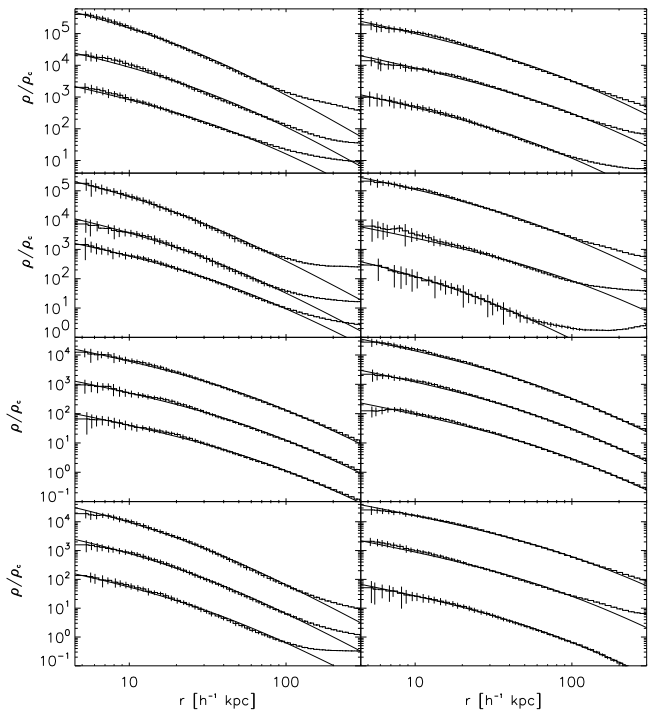,width=\hsize}}
      \caption{Density profiles for the four most massive satellites
               within halo~\#1 (left panel) and halo~\#2 (right panel)
               at $z=0.0$. For legibility, the profiles in each plot are shifted
               downwards by factors of 10 and
               correspond to (from top to bottom): \LCDM, \LWDM1, and \LWDM2.}
      \label{SatProfile}
    \end{figure}

This definition of the scale radius is more meaningful for the cases 
where we can not fit the data sufficiently well to the NFW profile as 
given by Eq.~(\ref{NFWfit}). This occurs mainly when the halo has undergone a
recent major merger. Results can be seen in Fig.~\ref{DensProfile1} and
Fig.~\ref{DensProfile2} which show the density profiles along with our
best fit NFW profiles for halo~\#1 and~\#2, respectively.
The error bars indicated by vertical lines for each bin in
the density profile figures (Fig.~\ref{DensProfile1}, 
Fig.~\ref{DensProfile2}, and Fig.~\ref{SatProfile}) are simply
Poissonian error bars. 

These figures should be viewed with Fig.~\ref{MassHistory} in mind; we
know that for halo~\#1 the first major merger happened between
redshift $z=1.5$ and $z=1.0$ and that is reflected by a steeper slope
of the density profile in the inner regions than predicted by the best
fit NFW model. The same situation can be observed from redshift
$z=0.4$ onwards where the profile starts to steepen albeit less
markedly, while the mass history indicates heavy accretion activity
(or minor merger events). A corresponding phenomenon is found for
halo~\#2 in Fig.~\ref{DensProfile2}. For this latter, the two
significant mergers occur between redshifts $z=0.5$ and $z=0.4$ and
$z=0.2$ and $z=0.1$, causing a steepening of the inner density profile
in all three (CDM as well as WDMs0 simulations.  We also tried to
fit our data to the even steeper profile proposed by Moore~\ea (1999a)
having an asymptotic slope $\rho(r) \propto r^{-1.5}$, but we were not
able to obtain sensible $\chi^2$ values.

\begin{figure}
      \centerline{\psfig{file=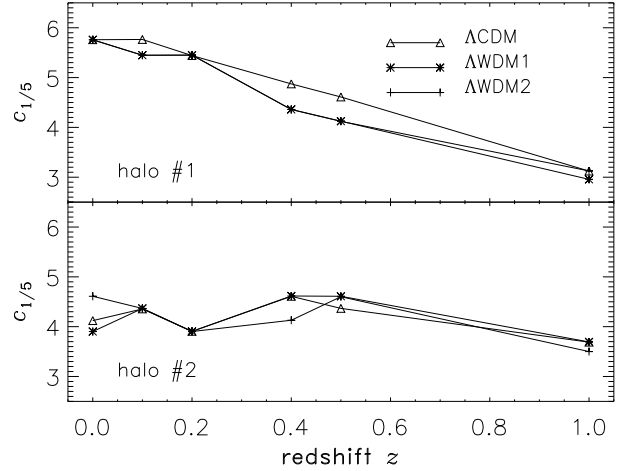,width=\hsize}}
      \caption{Redshift evolution of the concentration parameter
               \c15\ for the two most massive halos~\#1 and~\#2.}
      \label{cHistory}
    \end{figure}

However, in Fig.~\ref{cHistory} we plot the concentration parameter
\c15\ against redshift and do not find any abrupt changes or jumps for
the merger events identified in Fig.~\ref{MassHistory}. The only
noticeable trend is a nearly constant concentration parameter
\c15\ for halo~\#2. We stress that the fits of the density profiles
were only performed up to the virial radius of the halos (i.e. the
radius where $\rho(r)$ drops below 200 $\rho_{\rm crit}$). For
halo~\#2 we notice a moderate overshoot in the density profile at
large distances from the centre, and at nearly all redshifts.  This is
due to the fact that the progenitors of this halo all have important
masses and the centre of the profile is located at the centre of the
most massive one, therefore the most distant progenitor will create an
over-density which is not smeared out by the spherical averaging
process.

Finally, we point out that there are no obvious different trends
in the three different models: density profiles for halos in 
CDM and both WDM models all show the same behavior, with concentrations
 being only
marginally smaller for WDM halos than for corresponding CDM ones.

As already mentioned by other groups (Colin, Avila-Reese~\&
Valenzuela 2000 and Bode, Ostriker~\& Turok 2000) one should expect
the effect of a loss of concentration to be more pronounced for
satellite sub-halos orbiting in the host halos~\#1 and~\#2.  To verify
this, we identified the four most massive satellites ($M_{\rm
satellite} \approx (1 - 10) \cdot 10^{12}$\hMsun which roughly agrees
with 1000 -- 10000 particles per satellite) in both halos at redshifts
$z=0.0$ and $z=1.0$ and fitted their density profiles again with the
NFW profile given by Eq.~(\ref{NFWfit}). Even though the data is more
noisy for these halos, it is rather well described by such a profile,
as can be seen from Fig.~\ref{SatProfile}. 
We notice that the Poissonian error bars are systematically larger 
for halo~\#2 satellites in our \LWDM2\ simulation, due to a greater 
sensitivity to the tidal perturbations induced by the very recent merger.
These satellites are also slightly less massive than in the other 
simulations. It should be noted that,
as these satellites are embedded in the host halo, one cannot, in all
cases, define their virial radius via $\rho(r_{\rm vir})/\rho_c =
200$.  Their profiles sometimes flatten (or rise) and we therefore
manually truncated the halos in these cases. We then fitted their
density profiles up to either the virial radius or the point where the
profile flattened. We also computed the mean concentration \c15\
(averaged over the four satellites). Values for these quantities are
shown in Table~\ref{Cparameter}. This time, one can see a trend for
WDM sub-halos to have smaller concentrations than their CDM
counterparts. This holds at least for the more relaxed structure of
halo~\#1 whereas the overall concentrations for the satellites
orbiting in halo~\#2 are generally smaller because they are still in
the process of merging, and the differences between the three dark
matter models are therefore less pronounced.  An estimate of the
robustness of these results can be obtained by comparing the NFW fits
to the measured profiles in Fig.~\ref{SatProfile} where we only show
the fits to the satellites found in halo~\#1 and halo~\#2 at redshift
$z=0.0$. The concentration parameters are indeed well determined.

\begin{table}
\caption{Average concentration parameters \c15\ for satellites.}
\label{Cparameter}
\begin{tabular}{lcccc}
        & \multicolumn{2}{c}{halo\#1} & \multicolumn{2}{c}{halo\#2} \\ \hline
	& $z=0.0$ & $z=1.0$ & $z=0.0$ & $z=1.0$ \\ \hline \hline
\LCDM	& 7.065   & 3.783  & 4.360   & 3.603   \\
\LWDM1	& 5.967   & 3.004  & 3.992   & 3.517   \\
\LWDM2	& 4.475   & 3.376  & 4.145   & 3.296   \\
\end{tabular}
\end{table}

\subsection{Shapes of Halos}
The last section strongly argued in favour of a universal density
profile common to CDM and WDM halos; even satellite halos were found
to be well fitted by NFW profiles. However, there is still room for
subtle differences in the shapes of the halos as the computation of
density profiles implies redistributing particles inside spheres.

   \begin{figure}
      \centerline{\psfig{file=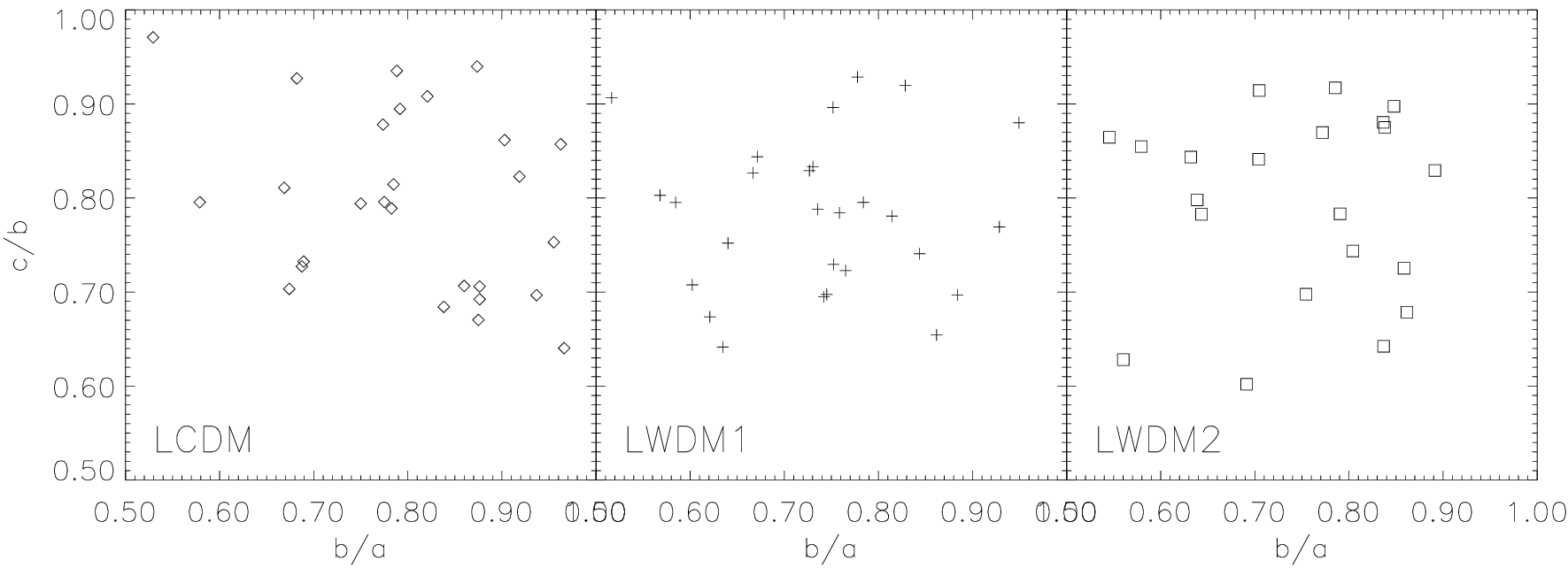,width=\hsize}}
      \caption{Axis ratios ($a>b>c$) for halos more massive
               than $3.5 \cdot 10^{12}$\hMsun\ (5000 particles).}
      \label{root}
    \end{figure}

Therefore, we have measured the triaxiality of all halos containing more than
5000~particles ($M>3.5 \cdot 10^{12}$\hMsun) using the usual
definition:

\begin{equation}
 T = \displaystyle \frac{a^2 - b^2}{a^2 - c^2}
\end{equation}

\noindent
where $a>b>c$ are the eigenvalues of the inertia tensor. Values
for a b and c can be read off of Fig.~\ref{root}, whereas Table~\ref{triaxial}
gives the value of T for the two most massive halos~\#1 and~\#2.

From Fig.~\ref{root} it is difficult again to pin down any obvious
difference between CDM and WDM simulated halos. Only
Table~\ref{triaxial} might indicate that the two most massive WDM
halos are marginally more spherical than their CDM counterparts (a
value of $T=1.0$ represents a prolate halo whereas $T=0.0$ means that
the halo is oblate).

\begin{table}
\caption{Triaxiality parameter $T$ for the two most massive halos~\#1 and~\#2.}
\label{triaxial}
\begin{tabular}{ll|l|l}
 halo    & \LCDM    &  \LWDM1    &  \LWDM2  \\ \hline \hline
 \#1:    & 0.97     &  0.94      &  0.88\\
 \#2:    & 0.90     &  0.70      &  0.70\\
 \end{tabular}
\end{table}

\subsection{Angular Momentum}
Another crucial issue in cosmological simulations of galaxies is
the angular momentum of simulated disks. 
There can only be two reasons why it is too low: either the halos 
in which the disks sit do not supply enough angular momentum to the baryons 
in the first place, or these baryons transfer too much of this angular 
momentum back to the dark matter when they cool to form the disk.
Sommer-Larsen and Dolgov found that the specific angular momentum 
of their disks was higher in WDM simulations than in CDM ones.
Admitting that their result holds, we can try to answer the question
as whether it is for one reason or the other, even though we do not have 
a cosmological hydrodynamic simulation.
For each individual halo containing more than 100 particles
($M>7 \cdot 10^{10}$\hMsun) we therefore computed the angular momentum
according to

\begin{equation} \label{angularmomentum}
 \vec{J} = \sum_{i=1}^{N} m_i \vec{r}_i \times \vec{v}_i \ .
\end{equation}

\noindent
This value is used to construct the dimensionless spin
parameter:

\begin{equation}
 \lambda = J \sqrt{|E|} / (G M^{5/2}) \ .
\end{equation}

   \begin{figure}
      \centerline{\psfig{file=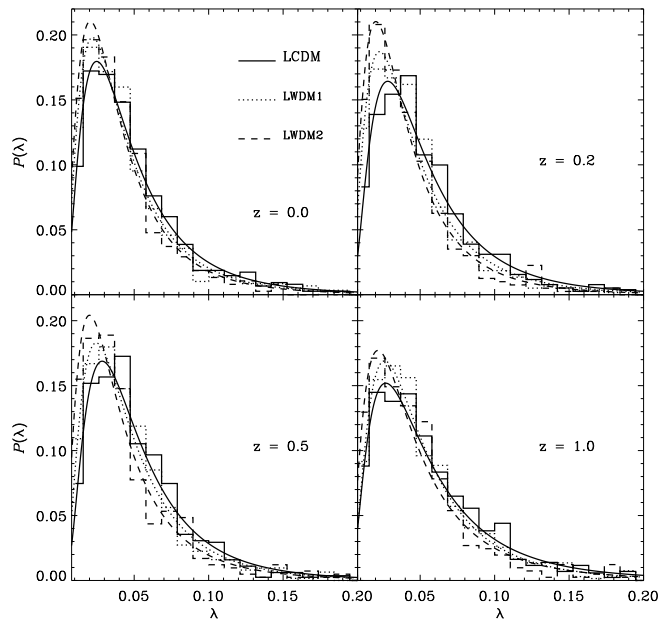,width=\hsize}}
      \caption{Spin parameter distribution in all three models
               for redshifts $z=0.0$, $z=0.2$, $z=0.5$, and $z=1.0$.
               Lines show fits obtained using the log-normal distribution 
		given by Eq.~(\ref{lognormal}).}
      \label{SpinFit}
    \end{figure}
\noindent
It has been pointed out before by several authors that the
distribution of this parameter follows a log-normal distribution
showing little evolution with redshift as well as little sensitivity
to the cosmological model (e.g. Frenk~\ea 1988; Warren~\ea 1992;
Cole~\& Lacey 1996; Gardner 2000, Maller, Dekel~\& Somerville 2001):

\begin{equation} \label{lognormal}
 n(\lambda) = \displaystyle \frac{1}{\lambda \sqrt{2\pi} \ \sigma_0}
              \exp \left( {-\frac{\ln^2 (\lambda/\lambda_0)}{2 \sigma_0^2}} \right)
\end{equation}

\begin{table*}
\caption{Redshift evolution of $\lambda_0$ when fitting
         spin parameter distribution to Eq.~(\ref{lognormal}).}
\label{SpinParam}
\begin{tabular}{lcccccccc}
       & \multicolumn{2}{c}{$z=0.0$}  & 
         \multicolumn{2}{c}{$z=0.2$}  & 
         \multicolumn{2}{c}{$z=0.5$}  & 
         \multicolumn{2}{c}{$z=1.0$} \\ \hline \hline
       & $\lambda_0$ & $\sigma_0$     & 
         $\lambda_0$ & $\sigma_0$     & 
         $\lambda_0$ & $\sigma_0$     & 
         $\lambda_0$ & $\sigma_0$  \\ \hline
\LCDM  & 0.040 & 0.693  & 0.054 & 0.681  & 0.069 & 0.658 & 0.093 & 0.682 \\
\LWDM1 & 0.036 & 0.707  & 0.046 & 0.717  & 0.061 & 0.690 & 0.086 & 0.703 \\
\LWDM2 & 0.034 & 0.716  & 0.042 & 0.723  & 0.052 & 0.742 & 0.082 & 0.768 \\

\end{tabular}
\end{table*}

\noindent
The first step is to check if this also holds for WDM simulations.
In Fig.~\ref{SpinFit} we show the fits of our numerical data using
Eq.~(\ref{lognormal}) at $z=0.0$, $z=0.2$, $z=0.5$, and $z=1.0$. This
figure is supplemented with the best fit values for the parameters
$\lambda_0$ and $\sigma_0$ in Table~\ref{SpinParam}.  These results show 
that there is no
clear correlation of the spin parameter with the dark matter particle mass, 
although one might argue for a marginal tendency of the halos to have 
slightly lower spin parameters in the WDM models.

\begin{figure}
      \centerline{\psfig{file=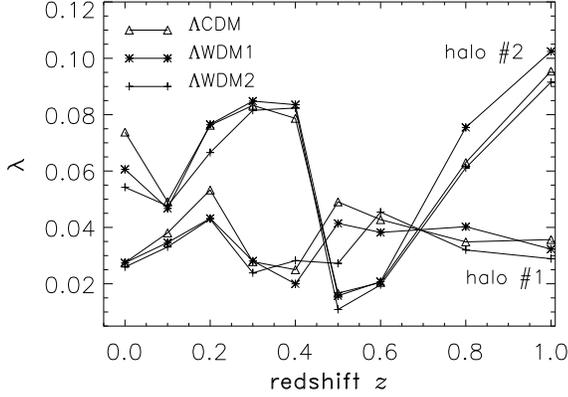,width=\hsize}}
      \caption{Spin parameter evolution for the two halos~\#1 and~\#2.}
      \label{SpinHistory}
    \end{figure}

One can also wonder what effect a recent merger event has on the spin
parameter. To answer this question we followed both spin histories
of massive halos~\#1 and~\#2, and plotted the results in
Fig.~\ref{SpinHistory}. There is no accidental evolution for halo~\#1
which mainly built up its mass via a steady accretion flow of material
(cf. Fig.~\ref{MassHistory}). But the situation looks different for
halo~\#2. For this latter, one can clearly spot the major merger happening
between $z=0.5$ and $z=0.4$ and possibly guess that another significant merger 
has happened around $z=0.1$.  But again, there is little (if any)
difference between \LCDM\ and the two \LWDM\ models. Changes
in angular momentum occurred in a similar fashion in all three models.

\begin{figure}
      \centerline{\psfig{file=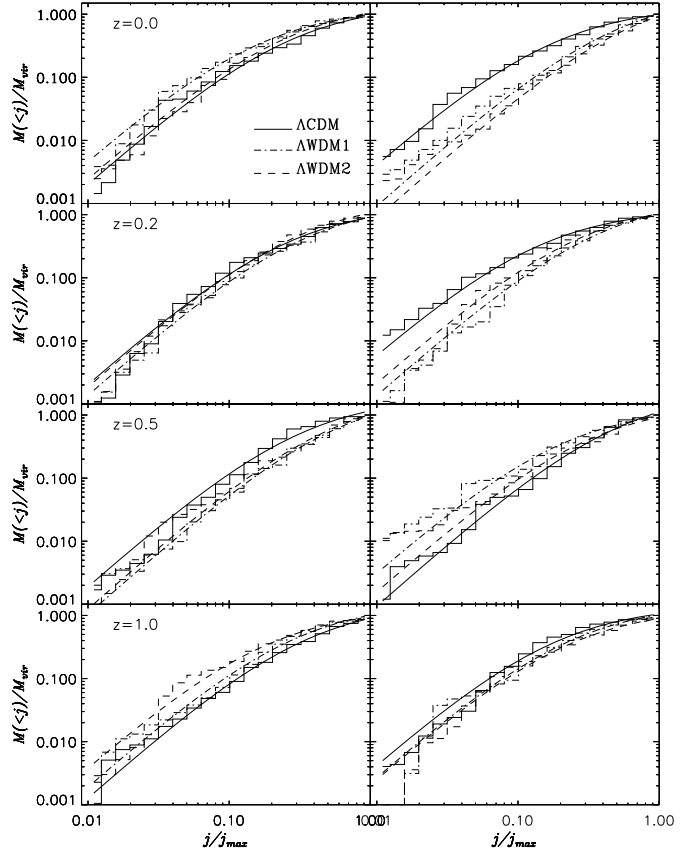,width=\hsize}}
      \caption{Mass distribution of specific angular momentum for the two
               halos~\#1 and~\#2 at various redshifts.}
      \label{MJ}
    \end{figure}

Inspired by the recent results from Bullock~\ea (2000) regarding a
universal angular momentum profile we also calculated the mass
distribution of angular momentum for the two halos~\#1 and~\#2.  This
was done by computing the angular momenta $\vec{J}$ of logarithmically
spaced shells ({\em i.e.} the same ones that were used for the density
profile). The specific angular momenta were obtained by simply
dividing the mass enclosed within each shell so that $j(r) =
|\vec{J}(r)|/M(r)$. Finally, the shells were sorted in order of
increasing $j$ and we looped over all possible $j$s from $j_{\rm
min}$ to $j_{\rm max}$, counting the cumulative mass in shells with
specific angular momentum less than $j$. The results can be seen in
Fig.~\ref{MJ} where we also present the redshift evolution of this
mass distribution of angular momentum for both halos.

The first thing we need to mention is that we were not able to fit the
profile suggested by Bullock~\ea (2000) to our data. Instead we
find that the data is better described by the following formula:

\begin{equation}
 M(<j) = \mu j^\alpha / (j_0 + j)^\alpha
\end{equation}

\noindent
with $\alpha \approx 2$. Those fits are presented as curves in
Fig.~\ref{MJ} whereas the data is given by the histograms.  But we
need to stress that our $M(<j)$ derivation slightly differs from
Bullock~\ea as we used spherical shells in contrast to their
construction of mass elements. Moreover, the power index $\alpha$
mainly affects the ''central'' parts of the distribution where $j$ is
less than about 6-7\% of $j_{\rm max}$.

However, it is clear from this figure that these distributions are
fairly similar, especially at redshift $z=0$, and that there is no
trend of a different time evolution between WDM and CDM halos.

As previously mentioned, Sommer-Larsen and Dolgov (2000) found that
galactic disks had higher specific angular momenta in their tree-SPH
WDM simulations than in their CDM simulations.  When one considers the
results presented above, it seems fairly safe to claim that this is
entirely due to a different transfer of angular momentum between gas
particles and dark matter sub-structures in their simulations: the
only noticeable difference in dark matter properties between WDM and
CDM being the suppression of these sub-structure in relaxed WDM
halos.

\section{Conclusions} \label{Conclusions}

We have run and analysed two \LWDM\ $N$-body simulations with realistic
warmon masses (0.5 and 1.0 keV) and compared them to an identical
\LCDM\ simulation. By identical, we mean that the initial conditions
in all the simulations were the same, except for the cut-off of the
power spectrum of the density field on small scales, characteristic of
warm dark matter particles typical free-streaming scales.  More
specifically, we have focussed on the detailed analysis of two
(cluster-sized) halos with mass $M \approx 9 \cdot 10^{13}$\hMsun, one
of which was found to be well relaxed, and the other still in the
process of virializing. In all the simulations, we have studied the
properties of these quite massive halos, but also of the substructure
they contained (mainly galaxy-sized objects).  We found that:

\begin{itemize}

 \item there are fewer sub-halos in WDM host halos when the mass of the
       dark particle drops.  This is due to the fact that in WDM
       scenarii satellites are more fluffy than in CDM scenarii
       (their concentration parameter is lower), which makes them more
       prone to disruption during the relaxation processes of the host
       halo.

 \item the formation times of small mass halos ($M \sim
	10^{10-11}$\hMsun) and their mode and sites of formation are
	different: a vast fraction of WDM small halos form in a
	top-down fashion by fragmentation of filaments at low
	redshifts, whereas in CDM simulations they all form in a
	bottom-up way through the merging of even smaller objects at
	high redshifts.  

\item   the average ages of dwarfs is sensibly higher in CDM models, 
	and these small galaxies are more void-filling than their WDM
	counterparts which track more closely the cosmic web.
 
 \item the merger and spin parameter histories are almost identical
       for CDM and WDM halos above this mass.

 \item density profiles of all our halos and sub-halos are well fit
       by NFW profiles.

 \item specific angular momentum profiles can be fitted by a universal
       profile as suggested by Bullock~\ea (2001) even though we found
       a steeper drop at the low $j$ end of that
       distribution. However, all curves are fairly similar in all
       models, to such an extent that the resolution of the angular
       momentum problem of disk galaxies in WDM simulations (which was
       claimed by Sommer-Larsen~\& Dolgov 2000) has to be entirely due
       to a different transfer between gas and dark matter
       substructures.
\end{itemize}

Comparing our work to previous analysis of a similar type, we find 
good agreement with the results of Colin et al. (2001), 
and we believe that the discrepancy between our work and that of Bode, 
Ostriker and Turok (2000) concerning concentration parameters can be explained 
by differences either in the warmon masses (they used lighter particles in 
several of their simulations) or in the data analysis. More specifically, 
we used \c15\ as a measure of the concentration of our halos which 
systematically yields lower values than the usual NFW concentration parameter 
that they use (Avila-Reese~\ea 1999; Colin~\ea 2000).
However, to really settle the issue, and quantify possible 
differences in the cores of massive WDM halos and in the overall reduction 
of the total number of small halos certainly requires more simulations
and of higher resolution to be performed.
As a conclusion, we stress that even though we showed that WDM is not the 
ideal solution to the so-called CDM crisis, it is still worth exploring 
seriously, especially if more accurate data confirms that dwarf galaxy 
properties are in real conflict with CDM model predictions. 

\section*{Acknowledgments}
AK would like to thank Anatoly Klypin and Andrey Kravtsov for kindly
providing a copy of the ART code as well as valuable comments. He
furthermore thanks Stefan Gottl\"ober, Vladimir Avila-Reese,
and Octavio Valenzuela for stimulating discussions during his stay
at New Mexico State University in Las Cruces in spring 2000 where
this project was initiated. AM thanks Martin Beer for
providing a helping hand at all times. Finally, we are also 
indebted to the referee, Paul Bode for pointing out an error in 
the calculation of the warmon mass.



\begin{thebibliography}{}

\bibitem[avila99]{avila99}
        {Avila-Reese V., Firmani C., Klypin A., Kravtsov A.V.,
         \MNRAS{310}{527}{1999}}

\bibitem[avila00]{avila00}
        {Avila-Reese V., Colin P., Valenzuela O., D'Onghia E., Firmani C.,
         \astroph{0010525}}

\bibitem[bahcall97]{bahcall97}
        {Bahcall N.A., Fan X., Cen R., \ApJL{485}{53}{1997}}

\bibitem[bahcall99]{bahcall99}
        {Bahcall N.A., Ostriker J.P., Perlmutter S., Steinhardt P.J.,
         \Science{284}{1481}{1999}}

\bibitem[bardeen]{bardeen}
	{Bardeen J.M., Bond J.R., Kaiser N., Szalay A.S., \ApJ{304}{15}{1986}}

\bibitem[barkana]{barkana}
	{Barkana R., Haiman Z., Ostriker J.P., \astroph{0102304}}

\bibitem[balbi]{balbi}
	{Balbi A., et al., \astroph{0005124}}

\bibitem[benson]{benson}
	{Benson A., et al., \astroph{0103092}}

\bibitem[bento]{bento}
	{Bento M.C., et al. Phys. Rev. D {\bf 62}, (2000)}

\bibitem[binney]{binney}
	{Binney J., Bissantz N., Gerhardt O., \ApJL{537}{99}{2000}}

\bibitem[blok]{blok01}
	{de Blok W.J.G., McGaugh S.S., Bosma A., Rubin V.C., \astroph{0103102}}

\bibitem[bode]{bode}
        {Bode P., Ostriker J.P., Turok N., \astroph{0010389}}

\bibitem[bullock]{bullock01}
	{Bullock J., Dekel A., Kolatt T.S., Kravtsov A.V., Klypin A.A., 
         Porciani C., Primack J.R., \astroph{0011001}}

\bibitem[chiu]{chiu}
        {Chiu W.A., Gnedin N.Y., Ostriker J.P., \astroph{0103359}}

\bibitem[cole96]{cole96}
        {Cole S., Lacey C., \MNRAS{281}{716}{1996}}

\bibitem[colin00]{colin00}
        {Colin P., Avila-Reese V., Valenzuela O., \ApJ{542}{622}{2000}}

\bibitem[colombi]{colombi}
        {Colombi S., Dodelson S., Widrow L.M., \ApJ{458}{1}{1996}}

\bibitem[debernardis]{debernardis}
	{de Bernardis P., et al., \Nature{404}{955}{2000}}

\bibitem[couchman91]{couchman91}
        {Couchman H.M.P, \ApJL{368}{23}{1991}}

\bibitem[davis]{davis}
        {Davis M., Efstathiou G., Frenk C.S., White S.D.M., \ApJ{292}{371}{1985}}

\bibitem[dolgov]{dolgov}
        {Dolgov A., Hansen S. H., {\tt hep-ph/0009083}}

\bibitem[dressler]{dressler}
        {Dressler A., Shectman S.A., \AJ{95}{985}{1988}}

\bibitem[eke96]{eke96}
        {Eke V.R., Cole S., Frenk C.S., \MNRAS{282}{263}{1996}}

\bibitem[eke98]{eke98}
        {Eke V.R., Cole S., Frenk C.S., Henry J.P., \MNRAS{298}{1145}{1998}}

\bibitem[frenk88]{frenk88}
        {Frenk C.S., White S.D.M., Davis M., Efstathiou G., \ApJ{327}{507}{1988}}

\bibitem[gardner00]{gardner00}
        {Gardner J., \astroph{0006342}}

\bibitem[hamilton]{hamiton2000}
        {Hamilton A., Tegmark M., \astroph{0008392}}

\bibitem[klypin97]{klypin97}
        {Klypin A.A., Holtzman J., \astroph{9712217}}

\bibitem[klypin99]{klypin99}
        {Klypin A.A., Kravtsov A.V., Valenzuela O., Prada F.,
         \ApJ{522}{82}{1999}}

\bibitem[knebe99]{knebe99}
        {Knebe A., M\"uller V., \AAA{341}{1}{1999}}

\bibitem[knebe00]{knebe00}
        {Knebe A., M\"uller V., \AAA{354}{761}{2000}}

\bibitem[maller]{maller}
        {Maller A.H., Dekel A., Somerville R.S., \astroph{0105168}}

\bibitem[meritt]{meritt}
	{Meritt D., Cruz F., \astroph{0101194}}

\bibitem[moore99a]{moore99a}
        {Moore B., Quinn T., Governato F., Stadel J., Lake G.,
         \MNRAS{310}{1147}{1999}}

\bibitem[moore99b]{moore99b}
        {Moore B., Ghigna S., Governato F., Lake G., Quinn T., Stadel J.,
         Tozzi P., \ApJL{524}{19}{1999}}

\bibitem[narayanan]{narayanan}
	{Narayanan V.K., Spergel D.N., Dav\'e R., Ma C.P., \astroph{0005095}}

\bibitem[perlmutter]{perlmutter}
        {Perlmutter S., et al., \ApJ{517}{565}{1999}}

\bibitem[pinkney]{pinkney}
        {Pinkney J., Roettiger K., Burn J.O., Bird C.M., \ApJ{104}{1}{1996}}

\bibitem[press]{press}
        {Press W.H., Schechter P., \ApJ{187}{425}{1974}}

\bibitem[riess]{riess}
        {Riess A.G., et al., \AJ{116}{1009}{1998}}

\bibitem[sahni]{sahni}
        {Sahni V., Colos P., Phys. Rep. {\bf 262}, 2 (1995)}

\bibitem[schmidt]{schmidt}
        {Schmidt B. P., et al.,  \ApJ{507}{46}{1998}}

\bibitem[seljak]{seljak}
        {Seljak U., Zaldarriaga M., \ApJ{469}{437}{1996}}

\bibitem[spergel]{spergel}
	{Spergel D.N., Steinhardt P.J., Phys. Rev. Lett. {\bf 84}, 3760 (2000)}

\bibitem[warren92]{warren92}
        {Warren M.S., Quinn P.J., Salmon J.K., Zurek W.H.,
         \ApJ{399}{405}{1992}}

\end{thebibliography}
\end{document}